\documentclass[%
reprint,
amsmath,amssymb,
aps,prd,
showkeys
]{revtex4-2}

\usepackage{graphicx}
\usepackage{dcolumn}
\usepackage{bm}
\usepackage{hyperref}
\hypersetup{
    colorlinks,
    citecolor=red,
    filecolor=black,
    linkcolor=blue,
    urlcolor=black
}
\usepackage{nicematrix}
\usepackage{tikz}
\usepackage{orcidlink}

\usepackage[arrowdel]{physics}
\newcommand{\eps}{\varepsilon}

\DeclareMathOperator*{\argmin}{argmin}

\begin{document}

\title{Bogoliubov-Born-Green-Kirkwood-Yvon hierarchy for quantum error mitigation}
\date{\today}

\author{Theo Saporiti \orcidlink{0009-0008-9738-3402}}
\email{theo.saporiti@cea.fr}
\author{Oleg Kaikov \orcidlink{0000-0002-9473-7294}}
\author{Vasily Sazonov \orcidlink{0000-0002-8152-0221}}
\author{Mohamed Tamaazousti \orcidlink{0000-0002-3947-9069}}
\affiliation{Université Paris-Saclay, CEA, List, F-91120, Palaiseau, France}

\begin{abstract}\noindent
Mitigation of quantum errors is critical for current NISQ devices. In the present work, we address this task by treating the execution of quantum algorithms as the time evolution of an idealized physical system. We use knowledge of its physics to assist the mitigation of the quantum noise produced on the real device. In particular, the time evolution of the idealized system obeys a corresponding BBGKY hierarchy of equations. This is the basis for the novel error mitigation scheme that we propose. Specifically, we employ a subset of the BBGKY hierarchy as supplementary constraints in the ZNE method for error mitigation. We ensure that the computational cost of the scheme scales polynomially with the system size. We test our method on digital quantum simulations of the lattice Schwinger model under noise levels mimicking realistic quantum hardware. We demonstrate that our scheme systematically improves the error mitigation for the measurements of the particle number and the charge within this system. Relative to ZNE we obtain an average reduction of the error by $(18.2 \pm 0.5)\%$ and $(52.8 \pm 6.3)\%$ for the respective above observables. We propose further applications of the BBGKY hierarchy for quantum error mitigation.
\end{abstract}


\maketitle

\section{Introduction}

The coupling of quantum computers to their surrounding environment is unavoidable, hence efficient methods to reduce quantum noise are highly demanded. While the general theory of quantum error correction offers a framework to achieve fully fault-tolerant computations \cite{ShorEC, Steane, Calderbank, Nielsen_Chuang_2010}, its required qubit overhead remains prohibitively high for today's quantum devices \cite{Chatterjee:2024kpt}. As an alternative to the currently challenging quantum error correction, quantum error mitigation (QEM) approaches were proposed \cite{PhysRevApplied.20.064027, PhysRevA.94.052325, PhysRevA.105.032620, Temme, Giurgica-Tiron:2020rcf, PracticalQEM, LearningBasedQEM, AIQEM, Cai:2020khs}. Despite their fundamental limitations \cite{FundamentalLimits, PhysRevLett.131.210602, FundamentalLimits2}, they remain the main available tools for the current noisy intermediate-scale quantum (NISQ) devices \cite{Pomarico:2025utk} and for the upcoming early phases of fault-tolerant computing \cite{PRXQuantum.3.010345, Zimboras:2025unr, Zhang:2025gyl}.

In this work, we empirically investigate how additional information provided by physics can improve the performance of QEM. The cornerstone idea of our approach is that the time-evolved state of any noiseless quantum system, at any time during the computation process, obeys a corresponding Schrödinger equation, so physical laws can be used to verify quantum computations. Unfortunately, this idea alone is of small practical use, since a full quantum state tomography of exponentially many measurements is required for the perfect knowledge of the state. To employ this concept in practice, one has to dramatically reduce the number of necessary measurements, for instance by symmetry verification \cite{Bonet-Monroig:2018mpi} or by using N-representability conditions \cite{Smart:2019hyn}.

We utilize the fact that the full dynamics of the idealized system of $N_\text{Q}$ qubits can be obtained from a quantum Bogoliubov-Born-Green-Kirkwood-Yvon (BBGKY) hierarchy \cite{Bogoliubov, Born:1946vqa, kirkwood, yvon1935theorie} of $4^{N_\text{Q}}$ equations. In classical computations, to avoid implementing an exponentially large system of coupled dynamical equations, one generally truncates the hierarchy by modeling the high-order correlators or by assuming their vanishing \cite{chari2016new, PhysRevB.93.174302, pavskauskas2012equilibration}. However, for strongly correlated systems or any general computational task, there are no (known) naturally small parameters justifying these truncations \cite{lacroix2016simplified, pavskauskas2012equilibration}. In quantum computations, truncations are no longer necessary, as one can directly measure any observable from the quantum device. One can then use corresponding equations from the BBGKY hierarchy to test the correctness of the measurements, hence of the quantum computations.

In this paper, we use the important fact that the amount of terms in all hierarchical equations is bounded by $\text{poly}(N_\text{Q})$, and by focusing on $\text{poly}(N_\text{Q})$-large subsets of the hierarchy, only a polynomial in $N_\text{Q}$ amount of additional classical resources is needed for the above-mentioned tests. We employ these supplementary informations from the BBGKY hierarchy to improve the zero-noise extrapolation (ZNE) method \cite{Temme, Li:2016vmf} in digital quantum simulations. Specifically, we formulate a novel QEM scheme and apply it to the Schwinger model \cite{PhysRev.128.2425} brought to quantum lattice simulations as a $\frac{1}{2}$-spin model in the particular implementation of \cite{schwinger}. The Schwinger model has been widely studied and used as a benchmark toy model for quantum computations, for instance in the recent works \cite{PhysRevD.105.014504,Pederiva:2022br,schwinger,Yamamoto:2022Qn,PhysRevD.105.094503,Ghim:2024pxe,Kaikov,DAnna:2024mmz}.

The paper is structured as follows. In section \ref{sec:bbgkyhierarchy} we derive and describe the
BBGKY hierarchy. Section \ref{sec:mitigationtechnique} is dedicated to our QEM method: in subsection \ref{subsec:znemethod} we briefly review the ZNE scheme, in subsection \ref{subsec:selectingbbgky} we describe how we select the BBGKY equations from the hierarchy, and in subsection \ref{subsec:ourmethod} we present our QEM technique. In section \ref{sec:resultsofmitigation} we apply our method to the lattice Schwinger model. Finally in section \ref{sec:conclusions} we conclude and discuss further potential applications of the method.

\section{The BBGKY hierarchy}\label{sec:bbgkyhierarchy}

We begin by deriving the BBGKY hierarchy and by discussing its properties.

Consider a quantum $\frac{1}{2}$-spin model composed of $N_\text{Q}$ qubits, each of which is labeled by an index $i \in \qty{1,\dots,N_\text{Q}} =: S$. Let $A \subseteq S$ represent a subsystem of the spin model, and let
\begin{equation}\label{eq:paulistring}
    \sigma(A, (\mu_i)_{i\in A}) := \prod_{i \in A} \sigma_i^{\mu_i}
\end{equation}
define a Pauli string, where $\mu_i \in \qty{1,2,3}$ and $\sigma_i^{\mu_i}$ is the Pauli operator acting on the $i$-th qubit in the $\mu_i$-th direction. Assume the model has an Hamiltonian of the form
\begin{equation}\label{eq:hamiltonian}
    H := \frac{1}{2} \sum_{i\in S} h_i^\mu \sigma_i^\mu + \frac{1}{4} \sum_{\substack{i,j\in S\\i < j}} V_{ij}^{\mu\nu} \sigma_i^\mu \sigma_j^\nu,
\end{equation}
where $h_i^\mu$ is the interaction term of the $i$-th spin in the $\mu$-th direction with an external magnetic field, $V_{ij}^{\mu\nu}$ is the interaction potential term among the $i$-th and $j$-th spins of respective $\mu$-th and $\nu$-th directions, and where from now on Einstein's summation is implied. Moreover, here and throughout the work, we set $\hbar = c = 1$.

If one injects \eqref{eq:paulistring} and \eqref{eq:hamiltonian} into Ehrenfest's theorem
and computes all the commutators, one obtains \cite{cox}
\begin{equation}\label{eq:bbgky}
\begin{split}
\dv{t} \ev{\prod_{i \in A} \sigma_i^{\mu_i}} &= \sum_{\substack{i,j\in A\\i \neq j}} \frac{V_{ij}^{\mu_i\nu}}{2}\eps_{\mu_j \nu \lambda}\ev{\sigma_j^\lambda \prod_{k \in A \setminus \qty{i,j}} \sigma_k^{\mu_k}}\\
&+\, \sum_{i\in A} h_i^\lambda \eps_{\mu_i \lambda \nu}\ev{\sigma_i^\nu \prod_{j \in A \setminus \qty{i}} \sigma_j^{\mu_j}}\\
&+\, \sum_{\substack{i \in A\\j\notin A}} \frac{V_{ij}^{\mu\nu}}{2}\eps_{\mu_i \mu \lambda}\ev{\sigma_i^\lambda \sigma_j^\nu \prod_{k\in A\setminus \qty{i}} \sigma_k^{\mu_k}},
\end{split}
\end{equation}
where $\eps_{\mu\nu\lambda}$ is the three-dimensional Levi-Civita symbol. We call \eqref{eq:bbgky} the BBGKY equation of the $\sigma(A, (\mu_i)_{i\in A})$ Pauli string. This is because, if one considers all the Pauli strings of all possible directions (namely all partitions $A \subseteq S$ of all possible $(\mu_i)_{i\in A}$) then, by computing all their associated BBGKY equations \eqref{eq:bbgky}, the complete exponentially large BBGKY-like hierarchy is generated. More precisely, for a specific Pauli string of length $\abs{A} = n \leq N_\text{Q}$, the time derivative of that $n$-point correlator is determined by a linear combination of $(n-1)$-point, $n$-point and $(n+1)$-point correlators, respectively found in the first, second and third summations of \eqref{eq:bbgky}, all selected according to the values $h_i^\mu$ and $V_{ij}^{\mu\nu}$ in \eqref{eq:hamiltonian}. The right-hand side (RHS) of \eqref{eq:bbgky} contains up to $9n(n-1)$-many $(n-1)$-point correlators, up to $9n$-many $n$-point correlators, and up to $27n(N_\text{Q}-n)$-many $(n+1)$-point correlators. Importantly, this implies that the amount of correlators in the RHS of \eqref{eq:bbgky} is polynomial in $n$ and $N_\text{Q}$, and bounded by $81 N_\text{Q}^2/8$.

As a last remark, note that \eqref{eq:hamiltonian} is constructed out of one-site and two-site interaction terms. A more general Hamiltonian containing higher-order Pauli strings would generate different BBGKY equations, with more terms in their RHS. However, this would not spoil the applicability of our method, as the resulting \eqref{eq:bbgky} equations would remain linear combinations of expectation values. In subsection \ref{subsec:ourmethod} we explain why this property is fundamental for the formulation of our mitigation scheme.

\section{Mitigation technique}\label{sec:mitigationtechnique}

In this section, we provide a short review of ZNE in the context of time evolution. Then we present our method, a BBGKY-improved ZNE scheme, which we formulate as a post-processing linear least-squares (LLS) optimization procedure. Our method incorporates a $\text{poly}(N_\text{Q})$-large subset of the BBGKY hierarchy evaluated across all time points. We consider the list of $l$ Pauli strings $\qty{Q_q}_{q\in\qty{0,\dots,l-1}}$, where for brevity $Q_q$ is defined as in \eqref{eq:paulistring}. Our goal is to mitigate the corresponding measurements obtained from a realistic quantum device. We focus solely on Pauli strings because they form an operator basis for the observables of the system.

\subsection{The ZNE method}\label{subsec:znemethod}
By Trotterization, the evolution time $T$ is discretized into $N$ slices of duration $\Delta t := T/N$, and evolution steps are obtained thanks to a Suzuki-Trotter decomposition scheme of order $d_\text{ST}$ \cite{82edc856-4d85-3b98-9b0d-ad55bb9315f6, Berry:2005yrf, Hatano:2005gh, Magnus:1954zz}. Then, different realizations of the quantum circuit implementing the time evolution are generated, each of them containing local unitary foldings \cite{Giurgica-Tiron:2020rcf} with a frequency of $\eta \geq 0$ insertions per step. Under the assumption that unitary foldings are affected by the same kind of noise as regular evolution steps, this implies an error level at the $s$-th step $s\in\qty{0,\dots,N}$ relative to the original $\eta=0$ circuit of
\begin{equation}\label{eq:errorlevel}
    1 \leq \eps_{s\eta} := \frac{s + 2 \left\lfloor \eta s \right\rfloor}{s} \xrightarrow{s \to \infty} 2\eta+1.
\end{equation}
Performing the above $m$ times with different noise levels $\eta \in (\eta_1,\dots,\eta_m) =: \va{\eta}$, at each time point $t_s:=s\Delta t$, we end up with an experimentally measured set of data points $(\eps_{s\eta},\ev{Q_q}_{s\eta})$, where $\ev{Q_q}_{s\eta}$ is the measurement of $Q_q$ at time point $t_s$, estimated with $N_\text{S}$ shots, under the $\eta$ noise level \footnote{We systematically add a small random $\order{1/\sqrt{N_\text{S}}}$ shift to $\eps_{s\eta}$ for reasons explained in appendix \ref{ap:random}}. For a given quantity at a given time point, these experimentally measured points can be fitted across the error levels with a chosen model function of $\eps$, leading to a zero noise $\eps \to 0$ extrapolated $\ev{Q^0_q}_s$ \cite{Giurgica-Tiron:2020rcf, Temme}. Common choices of fitting strategies include polynomials of degree $d \leq m - 1$, (poly-) exponential ansätze, and Richardson's extrapolation \cite{Giurgica-Tiron:2020rcf}.

\subsection{Selection of BBGKY equations}\label{subsec:selectingbbgky}

\begin{figure}
\centering
\includegraphics[width=0.3\textwidth]{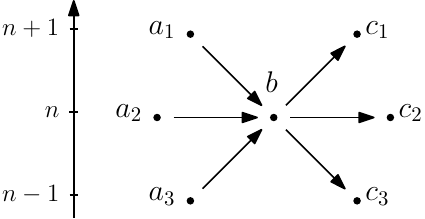}
\caption{In this diagram, $b$ is a Pauli string of interest, and $a_1, c_1$ are $(n+1)$-point correlators, $a_2, b, c_2$ are $n$-point correlators, and $a_3, c_3$ are $(n-1)$-point correlators. All $c_1,c_2,c_3$ are downstream connected to $b$, and all $a_1,a_2,a_3$ are upstream connected to $b$.}
\label{fig:connections}
\end{figure}

Given the set of expectation values $\qty{\ev{Q_q}}_{q\in\qty{0,\dots,l-1}}$, the physical knowledge provided by a BBGKY equation can help their mitigation if and only if its associated Pauli string is hierarchically connected to any of the $\qty{Q_q}_{q\in\qty{0,\dots,l-1}}$.

For a given generic $\sigma(A, (\mu_i)_{i\in A})$, the RHS of equation \eqref{eq:bbgky} provides all of the correlators $\sigma(B, (\nu_i)_{i\in B})$, with $B \subseteq S$ of directions $(\nu_i)_{i\in B}$, that are connected to its time evolution via the hierarchy. In that case, we say that the $\sigma(B, (\nu_i)_{i\in B})$ are downstream connected to $\sigma(A, (\mu_i)_{i\in A})$. We now want to determine the inverse, that is, given a Pauli string $\sigma(B, (\nu_i)_{i\in B})$, find all correlators $\sigma(A, (\mu_i)_{i\in A})$ generating $\sigma(B, (\nu_i)_{i\in B})$ in their RHS of \eqref{eq:bbgky}. In that case, we say that $\sigma(A, (\mu_i)_{i\in A})$ is upstream connected to $\sigma(B, (\nu_i)_{i\in B})$. Overall, we say that a correlator is connected to a Pauli string of interest if it is either downstream or upstream connected to that Pauli string.

To find all upstream connected correlators $\sigma(A, (\mu_i)_{i\in A})$ to a given $\sigma(B, (\nu_i)_{i\in B})$, there are only 3 possibilities: denoting $n_A = \abs{A}$ and $n_B = \abs{B}$, $\sigma(B, (\nu_i)_{i\in B})$ can appear in the BBGKY equation of $\sigma(A, (\mu_i)_{i\in A})$ either as an $(n_A+1)$-point, $n_A$-point or $(n_A-1)$-point correlator. The selection rules to find $\sigma(A, (\mu_i)_{i\in A})$ candidates in all three cases are derived in appendix \ref{ap:selection}. There we also show that, importantly, the upper bound for the amount of ansätze one has to check to pick up all the upstream connected $\sigma(A, (\mu_i)_{i\in A})$ is polynomial in $N_\text{Q}$. Figure \ref{fig:connections} schematically summarizes the two possible kinds of connections.

For our mitigation purposes, we select the BBGKY equations associated to the subset of the BBGKY hierarchy whose Pauli strings are connected to any of the $\qty{Q_q}_{q\in\qty{0,\dots,l-1}}$ by at most $r$ connections. The subset is obtained by iteratively computing over itself all of its downstream and upstream connected Pauli strings, as explained in appendix \ref{ap:selection}, for a total of $r$ iterations. This generates a list of $g$ BBGKY equations containing $\Lambda \geq l$ correlators, the actual amount of expectation values $\qty{\ev{Q_q}}_{q\in\qty{0,\dots,l-1, l, \dots,\Lambda-1}}$ that have to be measured.

\subsection{Our BBGKY-improved ZNE method}\label{subsec:ourmethod}

Our method aims at improving the ZNE-obtained measurements $\ev{Q^0_q}_s$ with better BBGKY-constrained $\ev{Q^\emptyset_q}_s$ extrapolations. As a first application of the above idea, in this work we use a least-squares polynomial (LSP) fitting strategy for ZNE. This is motivated by two reasons: generality of the noise and convex optimization. Regarding the first argument, we do not make any assumptions about the nature of the quantum noise, so that our method is independent of the particular employed quantum device. By the Weierstrass approximation theorem \cite{bernstein}, any underlying noise dependence of the measurements can be arbitrarily approached by a polynomial, up to an accepted error. Therefore, the choice of a LSP is sensible in the absence of any assumptions about the noise regime. It is then up to the user to select the degree $d$ of the LSP and accept its associated Weierstrass approximation error. For the second argument, the use of LSPs ensures that ZNE can be formulated as a LLS problem, rendering the fitting (hence the mitigation) a convex problem, for which a unique solution exists and is guaranteed to be found.

In this setting, where ZNE is a LLS minimization procedure, we want to incorporate the selected BBGKY equations as linear combinations of $\ev{Q^\emptyset_q}_s$. To do so, we approximate the time derivatives in \eqref{eq:bbgky} with derivatives of a Bernstein polynomial, fitting the $N+1$ extrapolated expectation values. We use (derivatives of) Bernstein polynomials because they are linear in the $\ev{Q^\emptyset_q}_s$ and because they uniformly converge to the functions (derivatives) they are fitting, with an error of the order $\order{1/N}$ \cite{floater}. Starting from the Bernstein polynomial \cite{bernstein}
\begin{equation}
    \textstyle\ev{Q^\emptyset_q}(t) := \displaystyle\sum_{s=0}^N \textstyle\ev{Q^\emptyset_q}_s\displaystyle b_{sN}\qty(\frac{t}{T}),
\end{equation}
constructed out of the $N+1$ Bernstein polynomial basis elements $b_{sN}(x)$ of degree $N$, with $x\in\qty[0,1]$ hence $t\in[0,T]$, then
\begin{equation}\label{eq:bernstein}
    \dv{t}\textstyle\ev{Q^\emptyset_q}\displaystyle(t) = \sum_{s=0}^{N} \textstyle\ev{Q^\emptyset_q}_s \displaystyle \beta_{sN}\qty(\frac{t}{T}),
\end{equation}
where we define
\begin{equation}
    \beta_{sN}(x) := \frac{1}{\Delta t}\begin{cases}
    -(1-x)^{N-1} & \text{if $s=0$}\\
    x^{N-1} & \text{if $s=N$}\\
        b_{s-1,N-1}(x) - b_{s,N-1}(x) & \text{otherwise}\\
    \end{cases}.
\end{equation}

Using the approximate time derivatives \eqref{eq:bernstein}, approximations of the BBGKY equations \eqref{eq:bbgky} can be expressed as linear combinations of $\ev{Q^\emptyset_q}_s$. Thereby, BBGKY equations can be cast inside the original ZNE LLS procedure as additional constraints of the minimization problem
\begin{equation}\label{eq:generalization}
    \argmin_{\va{a}}\frac{1}{2}\norm{\va{v}(\va{a})}^2 \quad \text{with} \quad \va{v}(\va{a}) := M \va{a} - \va{y},
\end{equation}
where $\va{a}$ will be defined later in \eqref{eq:avector} and where
\begin{equation}
    M := \mqty(M_{\va{\eta}}\\G) \;\;\;\;\;\text{with}\;\;\;\;\; M_{\va{\eta}} := \bigoplus_{q=0}^{\Lambda-1} \bigoplus_{s=1}^N \underbrace{\mqty(\eps_{s\eta_1}^d & \dots & 1\\
    \vdots & \ddots & \vdots\\
    \eps_{s\eta_m}^d & \dots & 1)}_{=: M_{s\va{\eta}}}.
\end{equation}
Here $M_{s\va{\eta}}$ is a Vandermonde-like matrix, $G$ is filled with $g(N+1)$ lines of appropriate entries encoding the corresponding BBGKY equations at every time point $t_s$, and the target vector $\va{y} := ((\va{y}_{01\va{\eta}}, \dots, \va{y}_{0N\va{\eta}}), \dots, (\va{y}_{\Lambda-1,1,\va{\eta}}, \dots, \va{y}_{\Lambda-1,N,\va{\eta}}), \va{g})$ sequentially groups all $s \geq 1$ experimental measurements $\va{y}_{qs\va{\eta}} := (\ev{Q_q}_{s\eta_1},\dots,\ev{Q_q}_{s\eta_m})$ together with the $g(N+1)$-long $\va{g}$ vector encoding the $s=0$ expectation values of the BBGKY equations. We do not mitigate any $\ev{Q^\emptyset_q}_0$ because they can be numerically computed with arbitrary precision at $t=0$, hence they are known a priori. Notice that by setting $g=0$ one recovers the original ZNE procedures, in which case $\ev{Q^\emptyset_q} = \ev{Q^0_q}$ and $M$ decouples into $N$ independent ZNEs, each minimizing the error on
\begin{equation}
    \ev{Q_q}_{s\eta} \approx \textstyle\ev{Q^\emptyset_q}_s\displaystyle + \sum_{\delta=1}^d a_{qs\delta} \eps_{s\eta}^\delta,
\end{equation}
where all LSP coefficients $a_{qs\delta}$ are packed into
\begin{equation}\label{eq:avector}
    \va{a} := ((\va{a}_{01}, \dots, \va{a}_{0N}), \dots, (\va{a}_{\Lambda-1,1}, \dots, \va{a}_{\Lambda-1,N})),
\end{equation}
with $\va{a}_{qs} := (a_{qsd}, \dots, a_{qs1}, \ev{Q^\emptyset_q}_s)$. In particular, $\ev{Q^\emptyset_q}_s$ can be extracted from $(\va{a})_p$ at index $p = p(q,s) = (d+1) + (s-1)(d+1) + q(d+1)N$. Finally, notice that $M$ is a rectangular matrix of polynomial size $\qty[mN\Lambda + g(N+1)] \times (d+1)N\Lambda$, and that $G$ couples together all extrapolated $\ev{Q^\emptyset_q}_s$ in two ways: across all time points, as in \eqref{eq:bernstein}, and according to their connections in the hierarchy, as in \eqref{eq:bbgky}. Figure \ref{fig:slices} graphically represents our method, and in appendix \ref{ap:Mmatrix} we give an example of an $M$ matrix.

\begin{figure}
\centering
\includegraphics[width=0.98\linewidth]{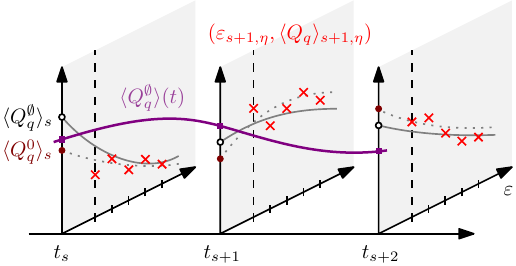}
\caption{Schematic depiction of our method: each slice of data (red crosses) is fitted with a LSP (dotted gray lines), producing a series of intermediate ZNEs (dark red points), which in turn are fitted with a Bernstein polynomial (purple line), giving access to time derivatives at each time point (purple squares). These are then used to improve the ZNEs thanks to the BBGKY equations, producing different LSPs (continuous gray lines) leading to BBGKY-improved ZNEs (black circles). The horizontal axis represents time, the vertical axes the expectation values, and the diagonal axes the error levels, where the dashed black lines represent the $\eps \geq 1$ bound.}
\label{fig:slices}
\end{figure}

\section{Results of the mitigation}\label{sec:resultsofmitigation}

We now briefly overview the lattice Schwinger model and the quantities we want to mitigate with our method. We then show and discuss the obtained numerical results.

\subsection{The lattice Schwinger model}\label{subsec:schwinger}

The Schwinger model \cite{PhysRev.128.2425} describes one-dimensional quantum electrodynamics. This continuous model can be brought to its lattice Hamiltonian formulation via Kogut-Susskind construction \cite{Kogut:1974ag}. Then, the original degrees of freedom can be recast into quantum $\frac{1}{2}$-spins \cite{Jordan:1928wi} with open boundary conditions \cite{schwinger}. We test our method on the latter, whose (dimensionless) Hamiltonian is
\begin{equation}
\begin{split}\label{eq:tavernelli}
    H &:= - \frac{m}{g}\sqrt{x} \sum_{i\in S} (-1)^i \sigma_i^3\\
    &+\sum_{i=1}^{N_\text{Q}-1} \qty(\frac{N_\text{Q}}{4}-\frac{1}{2}\left\lceil\frac{i-1}{2}\right\rceil +l_0(N_\text{Q}-i)) \sigma_i^3\\
    &+\frac{x}{2} \sum_{i=1}^{N_\text{Q}-1} \qty(\sigma_i^1\sigma_{i+1}^1 + \sigma_i^2\sigma_{i+1}^2)\\
    &+ \frac{1}{2}\sum_{\substack{i,j\in S\\i < j}} \qty(N_\text{Q}-j+\lambda) \sigma_i^3 \sigma_j^3,
\end{split}
\end{equation}
where $m/g$ is the lattice mass over coupling ratio, $x = \qty(N_\text{Q}/V)^2$ with $V$ the (dimensionless) lattice volume, $l_0$ the 
background electric field, $\lambda \gg 1$ a Lagrange multiplier to restrict simulations within the vanishing total charge sector, and a final constant term was disregarded. The Hamiltonian \eqref{eq:tavernelli} is in the appropriate form of \eqref{eq:hamiltonian}, and we are interested in the Pauli strings of the quantities
\begin{equation}
    Q := \frac{1}{2} \sum_{i\in S} \sigma^3_i \quad\text{and}\quad P := \frac{N_\text{Q}}{2} - \frac{1}{2} \sum_{i\in S} (-1)^i \sigma^3_i.
\end{equation}
These are, respectively, the electric charge operator and the particle number operator. Moreover $\comm{Q}{H} = 0$ and $\comm{P}{H} \neq 0$, meaning that they represent two distinct behaviors over which we can test our method: $\ev{P}$ will vary in time while $\ev{Q}$ will stay constant. In the following, we will often employ the abuses of notation $Q_q = Q,P$ with $q=Q,P$ to indicate the mitigation of the above linear combinations of Pauli strings.

\subsection{Numerical framework}\label{subsec:framework}

We assess the effectiveness of our method against ZNE with, respectively, the following $2$-norms
\begin{align}
    L^\emptyset_q &:= \sqrt{\Delta t\sum_{s=0}^{N} \qty[\ev{Q_q^\emptyset}_s - \ev{Q_q}(t_s)]^2},\\
    L^0_q &:= \sqrt{\Delta t\sum_{s=0}^{N} \qty[\ev{Q_q^0}_s - \ev{Q_q}(t_s)]^2},
\end{align}
which quantify the accumulated error of the extrapolations against the exact diagonalization (ED) evolution $\ev{Q_q}(t)$ over all time points $t_s$. All of our computations are performed in Qiskit 1.3 \cite{qiskit2024} within a simulated quantum device whose realistic noise model is generated in real time from the backend physical properties of the IBM Brisbane quantum processor \footnote{The simulations for $P$ were conducted on February 4th 2025 while those for $Q$ on February 5th 2025.}. In the following, we fix the parameters $N_\text{Q} = 4$, initial state $\ket{0101}$, $N_\text{S} = 10240$, $N = 20$, $T = 4$, $\va{\eta} = (0, 1, 1.5, 2)$, $d_\text{ST} = 1$, $d = 2$, $\lambda = 100$ and $V = 30$ \cite{schwinger}. Unless otherwise stated we fix $r=0$, while the remaining parameters $m/g$ and $l_0$ will vary throughout these simulations. We systematically check through idealized noiseless estimations ($N_\text{S} \to \infty$) of simulations that, for measurements of $\ev{Q}$ and $\ev{P}$, their total Trotter errors of order $\order{N(e\Delta t)^{d_\text{ST} + 1}}$, where $e$ is one unit of energy, are respectively smaller or equal to $1.9 \cdot 10^{-16}$ and $0.02$ with respect to ED evolutions. Therefore, as the Trotter error is negligible, only the shot noise of order $\order{1/\sqrt{N_\text{S}}}$ is considered for the computation of the errors associated to $L^\emptyset_q$ and $L^0_q$. These are respectively denoted $\Delta L^\emptyset_q$ and $\Delta L^0_q$, or respectively presented as $L^\emptyset_q \pm \Delta L^\emptyset_q$ and $L^0_q \pm \Delta L^0_q$, and they estimate the standard deviation of the two kinds of mitigations. Throughout the work, all standard deviation estimations are computed with the conventional propagation of errors formula, assuming independent variables and truncating at the leading order.

\subsection{Numerical results}\label{subsec:numericalresults}

\begin{figure}
\centering
\includegraphics[width=0.75\linewidth]{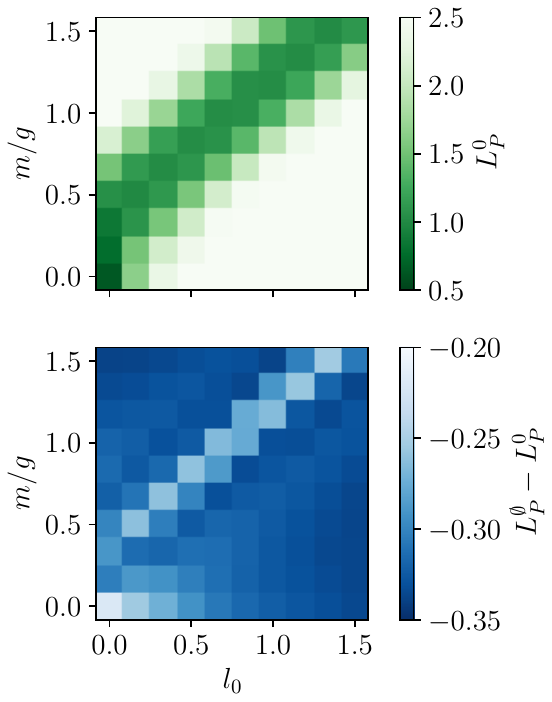}
\caption{Top panel: improvement of the ZNE mitigation for the particle number $\ev{P}$ with respect to the ED evolution. Bottom panel: advantage of our method compared to the top panel. In both panels $\Delta L^0_P \in [0.00891, 0.00920]$ and $\Delta L^\emptyset_P \in [0.00562, 0.00774]$.}
\label{fig:Pheat}
\end{figure}

\begin{figure}
\centering
\includegraphics[width=\linewidth]{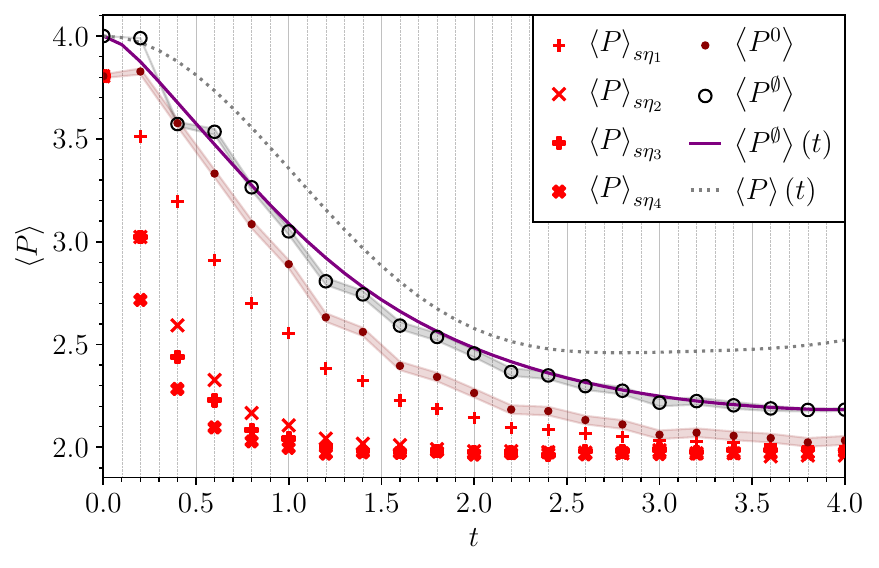}
\caption{Evolution of the particle number $\ev{P}$ in the $(l_0, m/g) = (0,0.15)$ regime. The red crosses are measurements at increasing noise levels (shot noise not shown, of average value $0.01$), the ZNEs are represented in dark red (estimated standard deviations shown as a band), the BBGKY-improved ZNEs are represented in black (estimated standard deviations shown as a band), the purple line is the Bernstein polynomial associated to the latter, and the dashed gray line is the ED evolution.}
\label{fig:Pdyn}
\end{figure}

Figure \ref{fig:Pheat} shows, for the $P$ observable, a parameter scan of $10 \times 10$ blocks over the region $(l_0, m/g) \in [0,1.5]^2$ where it is computed, in the top panel, the error of ZNE with respect to the ED dynamics and, in the bottom panel, the improvement of our method with respect to ZNE. The top panel gives us the scale of the ZNE error, and the bottom panel tells us by how much that error was reduced within our BBGKY-improved scheme. A systematic improvement over the entire parameter-region is manifest by the presence of only negative $L^\emptyset_P - L^0_P$ values and, taking average values as in table \ref{tbl:table}, it is approximately $(18.2 \pm 0.5)\%$. In all simulations $\Delta L^\emptyset_P < \Delta L^0_P$, meaning that our method reduces the estimated standard deviation of ZNE. The diagonal line in both panels can be explained by a system-dependent artifact caused by the alignment of the ED dynamics to the saturation, or flattening, of the measurements as in figure \ref{fig:Pdyn}.

Figure \ref{fig:Pdyn} shows the time evolution of the $(l_0,m/g) = (0,0.15)$ block of figure \ref{fig:Pheat}. We see that after ${\sim}15$ Trotter steps the measurements saturate, and recovering the original dynamics becomes challenging. Nevertheless, thanks to the additional BBGKY constraints, we see that the Bernstein polynomial correctly tries to match the time derivatives of the ED evolution.

\begin{figure}
\centering
\includegraphics[width=0.75\linewidth]{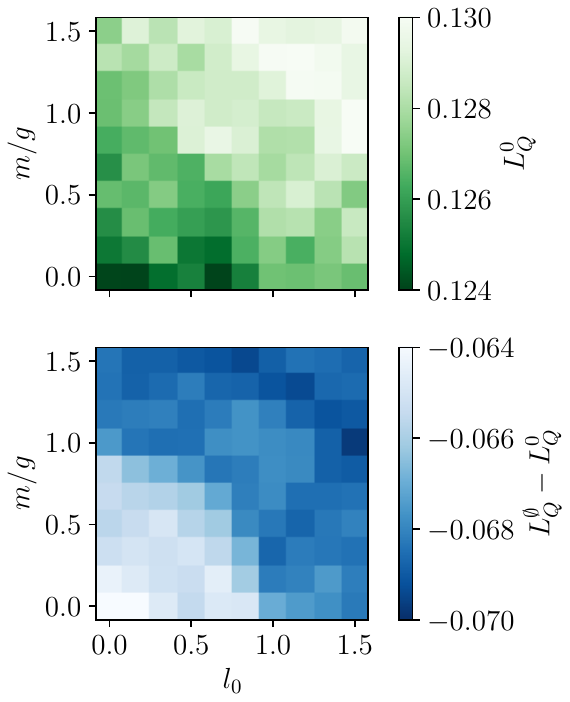}
\caption{Top panel: improvement of the ZNE mitigation for the electric charge $\ev{Q}$ with respect to the ED evolution. Bottom panel: advantage of our method compared to the top panel. In both panels $\Delta L^0_Q \in [0.00831, 0.00836]$ and $\Delta L^\emptyset_Q \in [0.00706, 0.00717]$.}
\label{fig:Qheat}
\end{figure}

Figure \ref{fig:Qheat} shows the contents of figure \ref{fig:Pheat} but for the $Q$ observable. Again, a systematic improvement over the entire parameter-region is observed with our method and, from table \ref{tbl:table}, we see that it is approximately $(52.8 \pm 6.3)\%$ with respect to ZNE. Once more the reduction $\Delta L^\emptyset_Q < \Delta L^0_Q$ of estimated standard deviation by our method is confirmed on all simulations. Here the bottom-left artifact region of the bottom panel can be explained by the small values of $(l_0, m/g)$, reducing the importance of the BBGKY equations in the LLS minimization.

\begin{figure}
\centering
\includegraphics[width=\linewidth]{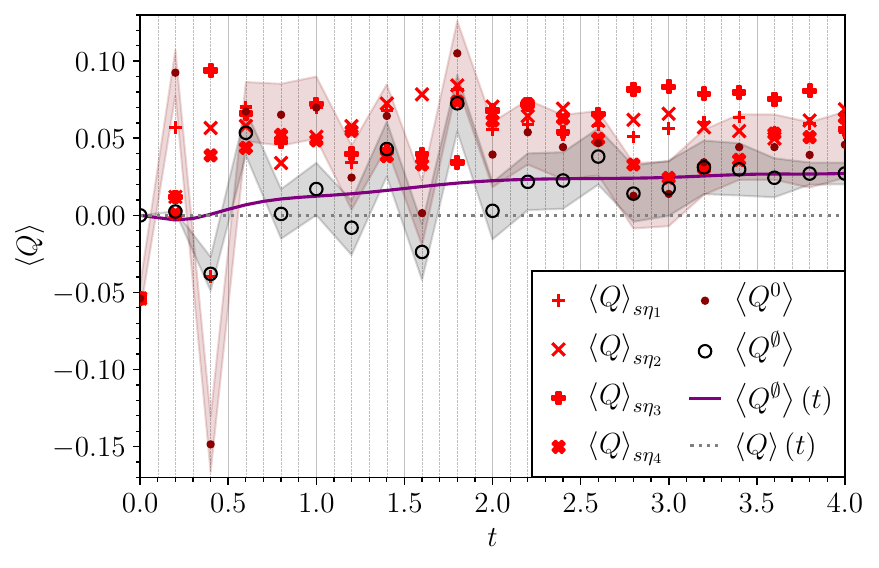}
\caption{Evolution of the electric charge $\ev{Q}$ in the $(l_0, m/g) = (0,0)$ regime. The same symbols and colors of figure \ref{fig:Pdyn} are employed, and the same average shot noise of the raw measurements is observed.}
\label{fig:Qdyn}
\end{figure}

Figure \ref{fig:Qdyn} shows the time evolution of the $(l_0,m/g) = (0,0)$ block of figure \ref{fig:Qheat}. Here, no saturation phenomenon occurs, because all measurements remain equally noisy. Nevertheless, thanks to the additional BBGKY constraints, we see again that the Bernstein polynomial correctly tries to match the null time derivative of the conserved quantity $Q$.

\renewcommand{\arraystretch}{1.4}
\begin{table}
\begin{ruledtabular}
\begin{tabular}{l|llll}
$q$ & $L^0_q \,\,\, \text{(ZNE)}$ &
$L^\emptyset_q \,\,\, \text{(Ours)}$ &
$L^0_q - L^\emptyset_q$ &
$1 - L^\emptyset_q/L^0_q$ \\[2pt]
\colrule
$P$ & $1.975 \pm 0.009$ & \textbf{1.659 $\pm$ 0.007} & $0.316 \pm 0.011$ & $(18.2 \pm 0.5)\%$\\
$Q$ & $0.128 \pm 0.008$ & \textbf{0.060 $\pm$ 0.007} & $0.068 \pm 0.011$ & $(52.8 \pm 6.3)\%$\\
\end{tabular}
\end{ruledtabular}
\caption{\label{tbl:table}%
First column: average errors of ZNE against ED. Second column: average errors of our method against ED. Third column: absolute average improvements of our method against ZNE. Fourth column: relative average improvements of our method against ZNE. The averages for the $q=P,Q$ observables are computed, respectively, from the data of figures \ref{fig:Pheat} and \ref{fig:Qheat}.
}
\end{table}

Table \ref{tbl:table} summarizes as averages the errors, the absolute and the relative improvements of the previous two parameter scans, and their associated standard deviation estimations, shown in figures \ref{fig:Pheat} and \ref{fig:Qheat}. Again, with our method, we see a systematic improvement of the two mitigations with respect to ZNE, for both the non-conserved $P$ and the conserved $Q$, although we observe a larger relative improvement in the mitigation of $\ev{Q}$. This is because noise concentrates around the ED evolution in figure \ref{fig:Qdyn} so, in the minimization of the LLS problem, the Bernstein polynomial is less penalized in deviating from the ZNE to match the null time derivative of $\ev{Q}$.

\begin{figure}
\centering
\includegraphics[width=0.7\linewidth]{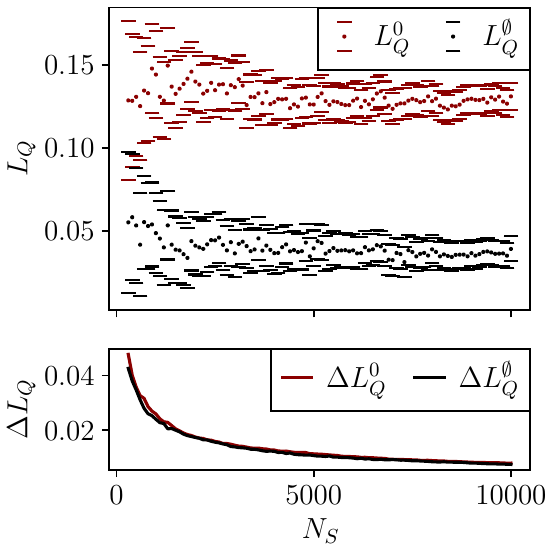}
\caption{Top panel: errors for the mitigation of the charge $\ev{Q}$, in the $(l_0, m/g) = (0,0)$ regime at different $N_\text{S}$, with ZNE (data represented in dark red) and our method (data represented in black). Bottom panel: magnitudes of the above panel's error bars.}
\label{fig:Qshot}
\end{figure}

Figure \ref{fig:Qshot} shows the error committed by ZNE and our BBGKY-informed method over the ED evolution of $\ev{Q}$ in the $(l_0,m/g) = (0,0)$ regime, using measurements estimated with an increasing number $N_\text{S}$ of shots. The top panel displays a comparable convergence of both mitigations to corresponding limiting errors as $N_\text{S}$ increases. Moreover, the bottom panel highlights the similarities in the magnitudes of the corresponding estimated standard deviations. Therefore, as our BBGKY-informed scheme behaves similarly to ZNE against statistical noise, the inherited stability and robustness of the LLS ZNE scheme are recovered in our method as well. Comparable results are obtained for the mitigation of $\ev{P}$ in the $(l_0,m/g) = (0,0)$ block.

\begin{figure}
\centering
\includegraphics[width=0.7\linewidth]{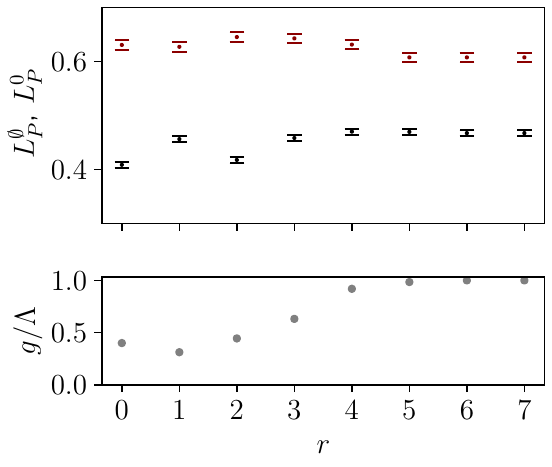}
\caption{Top panel: errors for the mitigation of the particle number $\ev{P}$, in the $(l_0, m/g) = (0,0)$ regime at different $r$, with ZNE (data represented in dark red) and our method (data represented in black). Bottom panel: determination of the hierarchical subset.}
\label{fig:Phie}
\end{figure}

We now study how the size of the selected subset of BBGKY equations affects the results of the mitigation. This is displayed in figure \ref{fig:Phie}, where the mitigation of $\ev{P}$ in the $(l_0,m/g) = (0,0)$ block is repeated for different maximal connections radii $r$. The top panel displays the errors of the ZNE and BBGKY-improved mitigations, while the bottom panel quantifies how many BBGKY equations $g$ cover the dynamics of the measured $\Lambda$ quantities.

A ratio of $g/\Lambda = 1$, which is reached at $r=6$ and remains so at $r=7$, means that the subset of equations is fully determined with respect to its unknowns. In the present case $\Lambda = 126 < 256 = 4^{N_\text{Q}}$, because the BBGKY hierarchy splits into 4 independent hierarchies of sizes 1, 1, 126 and 128. In particular, the two first hierarchies are composed, respectively, of solely the identity and $\sigma_1^3\sigma_2^3\sigma_3^3\sigma_4^3$, which are conserved quantities of \eqref{eq:tavernelli}. The third hierarchy involves $\{\sigma_i^3\}_{i\in S}$, which are required to build the $Q$ and $P$ observables.

We see that the inclusion of additional BBGKY constraints reduces the error with respect to ZNE immediately starting from $r = 0$, in accordance with figure \ref{fig:Pheat}. In our experiments with the Schwinger model, no clear pattern regarding the reduction or the increase of absolute errors can be stated for $r > 0$, neither in this $(l_0,m/g) = (0,0)$ simulation nor in other different points of the parameter scan. We conjecture that a similar behavior is valid for other systems and therefore the optimal regime of our method lies close to a small $r = 0$ hierarchical subset. This is because, given the ratio of the number of lines in the $M$ matrix distributed among ZNE and the BBGKY equations, the LLS problem prevents the Bernstein polynomial from deviating too much from the original ZNEs even as $g$ increases. Moreover, the number of time points $N$ may be too small for the Bernstein polynomial to benefit from its uniform convergence, leading to a systematic non-negligible additional approximation error from \eqref{eq:bernstein}. Comparable results and conclusions are observed for the mitigation of $\ev{Q}$ in the $(l_0,m/g) = (0,0)$ block.

\section{Conclusions}\label{sec:conclusions}

In this paper, we considered executions of quantum algorithms as time evolutions of idealized systems. Their noiseless dynamics are governed by physical laws, which can be employed to mitigate the quantum errors arising from their physical realizations. For this purpose, we derived a corresponding BBGKY-like hierarchy and selected a $\text{poly}(N_\text{Q})$-large subset of its equations. We proposed a novel QEM scheme encapsulating these supplementary physics-informed constraints into the ZNE procedure. A LLS problem is thereby obtained, whose required classical computational resources scale polynomially in $N_\text{Q}$. We numerically investigated the effectiveness of our method on digital quantum simulations, mimicking realistic quantum hardware noise, of the lattice Schwinger model.

By applying our method to the lattice Schwinger model, we assessed our BBGKY-informed QEM scheme against ZNE by comparing the mitigation of quantum observables to known ED evolutions. It was found that, in the considered regions of the parameter scan and under the selected input parameters, our method systematically improves the QEM of ZNE measurements. Moreover, with our method, it was found that the range of relative improvement of the error with respect to ZNE spans from $(18.2 \pm 0.5)\%$ to $(52.8 \pm 6.3)\%$, depending on whether the mitigated quantity is conserved or not. It was also found that our method exhibits the same stability and robustness properties of conventional ZNE. Finally, it was found that the maximal connections radius of the selected hierarchical subset should be close to $r=0$, as such a subset of BBGKY equations already guarantees a systematic improvement over ZNE.

Further expansion of this work includes considering Hamiltonians encoding more general multi-spin interactions, performing imaginary time evolution, and evolution with time-dependent Hamiltonians. The latter would pave the way to the mitigation of adiabatic time evolution. This would also allow the mitigation of quantum circuits, if implemented as Trotterized time evolutions of a time-dependent system. A more detailed analysis of the stability and robustness of our method could be investigated, and different non-linear fitting strategies other than LSPs could be implemented and tested. Finally, the supplementary physical knowledge of the BBGKY hierarchy could be used to not only help ZNE but other mitigation schemes as well, either on digital or analog quantum machines. This could be done either as an entirely post-processing procedure, or directly affecting the parameters of quantum computation, as in variational methods.

\begin{acknowledgments}
This work has received support from the French State managed by the National Research Agency under the France 2030 program with reference ANR-22-PNCQ-0002. We acknowledge the use of IBM Quantum services for this work. The views expressed are those of the authors, and do not reflect the official policy or position of IBM or the IBM Quantum team.
\end{acknowledgments}

\section*{Data Availability}
The data that support the findings of this article are openly available \cite{dataset}.

\appendix

\section{Upstream connected correlators}\label{ap:selection}
Here we derive an algorithm composed of three subroutines to obtain all $\sigma(A, (\mu_k)_{k\in A})$ correlators upstream connected to the target $\sigma(B, (\nu_k)_{k\in B})$. First, define the function $\bar{\eps}(\mu,\nu) := \lambda \abs{\eps_{\mu\nu\lambda}}$, where Einstein's summation is still implied, to be interpreted as that index such that $\eps_{\mu\nu\bar{\eps}(\mu,\nu)} \neq 0$. Let also $n_A = \abs{A}$ and $n_B = \abs{B}$.

For the $n_A = n_B$ case, we observe in the second summation of \eqref{eq:bbgky} that, for $\sigma(B, (\nu_k)_{k\in B})$ to appear in the expectation value, it must be $A=B$. Regarding the directions $(\mu_k)_{k\in A}$, selecting an $i\in A$ ($n_B$ choices) and letting $\mu_k = \nu_k$ for all $k \in A \setminus \qty{i}$, we see that if we want $\nu$ to pick up the desired $\nu = \nu_i$ value then, because of the Levi-Civita symbol, it must be $\mu_i \neq \nu_i$ (2 choices). Then $\lambda = \bar{\eps}(\mu_i,\nu_i)$, hence $\sigma(B, (\nu_i)_{i\in B})$ appears in the second summation only if
\begin{equation}
    h_i^{\bar{\eps}(\mu_i,\nu_i)} \neq 0.
\end{equation}
In total, $2n_B$ possible $\sigma(A, (\mu_k)_{k\in A})$ ansätze have to be checked against the magnetic field.

For the $n_A = n_B - 1$ case, we observe in the third summation of \eqref{eq:bbgky} that, for $\sigma(B, (\nu_k)_{k\in B})$ to appear in the expectation value, it must be $A=B\setminus\qty{j}$ for a selected $j\in B$ ($n_B$ choices). Regarding the directions $(\mu_k)_{k\in A}$, selecting a specific $i\in A$ ($n_B - 1$ choices) and letting $\mu_k = \nu_k$ for all $k\in A \setminus \qty{i}$, we see that if we want $\lambda$ to pick up the desired $\lambda = \nu_i$ value then, again because of the Levi-Civita symbol, it must be $\mu_i \neq \nu_i$ (2 choices). Then $\mu = \bar{\eps}(\mu_i,\nu_i)$, hence $\sigma(B, (\nu_k)_{k\in B})$ appears in the third summation only if
\begin{equation}
    V_{ij}^{\bar{\eps}(\mu_i,\nu_i)\nu_j} \neq 0.
\end{equation}
Notice that the index $\nu$ will necessarily pick $\nu = \nu_j$ because it is contracted independently. In total, $2n_B(n_B-1)$ possible $\sigma(A, (\mu_k)_{k\in A})$ ansätze have to be checked against the interaction potential.

For the $n_A = n_B + 1$ case, we observe in the first summation of \eqref{eq:bbgky} that, for $\sigma(B, (\nu_k)_{k\in B})$ to appear in the expectation value, it must be $A=B\cup\qty{i}$ for a selected $i\in S\setminus B$ ($N_\text{Q} - n_B$ choices). Regarding the directions $(\mu_k)_{k\in A}$, selecting a specific $j\in B$ ($n_B$ choices) and letting $\mu_k = \nu_k$ for all $k\in A \setminus \qty{i,j}$ and $\mu_i \in \qty{1,2,3}$ (3 choices), we see that if we want $\lambda$ to pick up the desired $\lambda = \nu_j$ value then, once again because of the Levi-Civita symbol, it must be $\mu_j \neq \nu_j$ (2 choices). Then $\nu = \bar{\eps}(\mu_j,\nu_j)$, hence $\sigma(B, (\nu_k)_{k\in B})$ appears in the first summation only if
\begin{equation}
    V_{ij}^{\mu_i\bar{\eps}(\mu_j,\nu_j)} \neq 0.
\end{equation}
In total, $6n_B(N_\text{Q} - n_B)$ possible $\sigma(A, (\mu_k)_{k\in A})$ ansätze have to be checked against the interaction potential.

To sum things up, there is a polynomial in $n_B$ and $N_\text{Q}$ amount of ansätze to check against the coefficients of \eqref{eq:hamiltonian}, and the number of checks is overall bounded by $9N_\text{Q}^2/4$.

\section{Random shift of error levels}\label{ap:random}
Every time \eqref{eq:errorlevel} is employed to measure a $(\eps_{s\eta},\ev{Q_q}_{s\eta})$ data point, a small random shift $\eps_{s\eta} \to \eps_{s\eta} + \chi$ is performed, where $\chi$ is a realization of the normally distributed random variable $X \sim \mathcal{N}(0, 1/N_\text{S})$. This is to avoid cases where two different noise levels $\eta_1 \neq \eta_2$ produce the same error level, for example $\eta_1 = 1$ and $\eta_2 = 1.5$ leading to $\eps_{1\eta_1} = \eps_{1\eta_2} = 3$, thereby introducing artifacts in the LSP fitting. The variance of $X$ is chosen to be $1/N_\text{S}$ to mimic the standard deviation
estimator when $N_\text{S}\to\infty$ and also because, in that limit, the $\chi$ shift shouldn't affect $\eps_{s\eta}$, as quantum errors had an infinite $N_\text{S}\to\infty$ amount of possibilities to arise.

\section{Example of an $M$ matrix}\label{ap:Mmatrix}
To illustrate the construction of $M$ with a simple yet non-trivial example, consider the mitigation of $l=\Lambda=2$ quantities after $N=2$ Trotter steps with the $g=2$ fictitious BBGKY equations
\begin{equation}
\dv{t}\ev{Q_0} = 0 \qquad\text{and}\qquad \dv{t}\ev{Q_1} = V\ev{Q_0},
\end{equation}
where $V \neq 0$ is constant and $Q_0,Q_1$ are, respectively, $n$-point and $(n+1)$-point correlators. Then, in this example, the $\va{v}$ vector is given by
\begin{widetext}
\,\\
\NiceMatrixOptions{cell-space-limits = 2.5pt}
\small
\begin{equation}
\va{v}(\va{a}) \, = \,\,\begin{pNiceMatrix}[r, margin, first-col]
\Block{3-1}{M_{1\va{\eta}} \quad\,} & \Block[borders={right, tikz=dashed}]{18-6}{}\Block[borders={top,bottom,left,tikz=dashed}]{6-6}{}\Block[borders={right,bottom,tikz=dotted}]{3-3}{}
\eps_{1\eta_1}^d    & \Cdots & 1      &                   &        &        &                   &        &        &                  &        &       \\
& \Vdots            & \Ddots & \Vdots &                   &        &        &                   &        &        &                  &        &       \\
& \eps_{1\eta_m}^d  & \Cdots & 1      &                   &        &        &                   &        &        &                  &        &       \\
\Block{3-1}{M_{2\va{\eta}} \quad\,} &                   &        &        &
\Block[borders={top,left,tikz=dotted}]{3-3}{}
                                        \eps_{2\eta_1}^d  & \Cdots & 1      &                   &        &        &                  &        &       \\
&                   &        &        & \Vdots            & \Ddots & \Vdots &                   &        &        &                  &        &       \\
&                   &        &        & \eps_{2\eta_m}^d  & \Cdots & 1      &                   &        &        &                  &        &       \\
\Block{3-1}{M_{1\va{\eta}} \quad\,} &                   &        &        &                   &        &        &
\Block[borders={top,right,tikz=dashed}]{6-6}{}\Block[borders={right,bottom,tikz=dotted}]{3-3}{}
                                                                            \eps_{1\eta_1}^d  & \Cdots & 1      &                  &        &       \\
&                   &        &        &                   &        &        & \Vdots            & \Ddots & \Vdots &                  &        &       \\
&                   &        &        &                   &        &        & \eps_{1\eta_m}^d  & \Cdots & 1      &                  &        &       \\
\Block{3-1}{M_{2\va{\eta}} \quad\,} &                   &        &        &                   &        &        &                   &        &        &
\Block[borders={top,left,tikz=dotted}]{3-3}{}
                                                                                                                  \eps_{2\eta_1}^d & \Cdots & 1     \\
&                   &        &        &                   &        &        &                   &        &        & \Vdots           & \Ddots & \Vdots\\
 & \Block[borders={bottom, tikz=solid}]{1-12}{}
                    &        &        &                   &        &        &                   &        &        & \eps_{2\eta_m}^d & \Cdots & 1     \\

\Block{6-1}{G \quad\,} & \Block[borders={right,tikz=dotted}]{6-3}{}
\Block[r]{1-3}{\beta_{12}(0)}&&&
\Block[r]{1-3}{\beta_{22}(0)}&&&
\Block[borders={right,tikz=dotted}]{6-3}{}
\Block[r]{1-3}{}&&&
&&\\

& \Block[r]{1-3}{\beta_{12}(\frac{1}{2})}&&&
\Block[r]{1-3}{\beta_{22}(\frac{1}{2})}&&&
&&&
\Block[r]{1-3}{}&&\\

& \Block[r]{1-3}{\beta_{12}(1)}&&&
\Block[r]{1-3}{\beta_{22}(1)}&&&
&&&
\Block[r]{1-3}{}&&\\

& \Block[borders={top, tikz=solid}]{1-12}{}

&&&&&&
\Block[r]{1-3}{\beta_{12}(0)}&&&
\Block[r]{1-3}{\beta_{22}(0)}&&\\

&\Block[r]{1-3}{-V}&&&&&&
\Block[r]{1-3}{\beta_{12}(\frac{1}{2})}&&&
\Block[r]{1-3}{\beta_{22}(\frac{1}{2})}&&\\

& &&&
\Block[r]{1-3}{-V}&&&
\Block[r]{1-3}{\beta_{12}(1)}&&&
\Block[r]{1-3}{\beta_{22}(1)}&&\\

\CodeAfter
    \OverBrace[yshift=5pt]{1-1}{1-6}{Q_0}
    \OverBrace[yshift=5pt]{1-7}{1-12}{Q_1}
    \OverBrace[yshift=5pt]{7-7}{7-9}{t_1}
    \OverBrace[yshift=5pt]{7-10}{7-12}{t_2}
    \UnderBrace[yshift=5pt]{6-1}{6-3}{t_1}
    \UnderBrace[yshift=5pt]{6-4}{6-6}{t_2}
    \SubMatrix{.}{1-1}{3-1}{\{}[xshift=-14mm]
    \SubMatrix{.}{4-1}{6-1}{\{}[xshift=-14mm]
    \SubMatrix{.}{7-1}{9-1}{\{}[xshift=-14mm]
    \SubMatrix{.}{10-1}{12-1}{\{}[xshift=-14mm]
    \SubMatrix{.}{13-1}{18-1}{\{}[xshift=-14mm]
    \UnderBrace[yshift=5pt]{1-1}{last-last}{M}
\end{pNiceMatrix}
\,\cdot\,\,
\begin{pNiceMatrix}[last-col]
a_{01d}&\Block{3-1}{\quad \va{a}_{01}}\\
\Vdots&\\
\Block[borders={bottom, tikz=dotted}]{1-1}{}\ev*{Q^\emptyset_0}_1&\\
a_{02d}&\Block{3-1}{\quad \va{a}_{02}}\\
\Vdots&\\
\Block[borders={bottom, tikz=dashed}]{1-1}{}\ev*{Q^\emptyset_0}_2&\\
a_{11d}&\Block{3-1}{\quad \va{a}_{11}}\\
\Vdots&\\
\Block[borders={bottom, tikz=dotted}]{1-1}{}\ev*{Q^\emptyset_1}_1&\\
a_{12d}&\Block{3-1}{\quad \va{a}_{12}}\\
\Vdots&\\
\ev*{Q^\emptyset_1}_2&
\CodeAfter
    \SubMatrix{.}{1-1}{3-last}{\}}[xshift=2.5mm]
    \SubMatrix{.}{4-1}{6-last}{\}}[xshift=2.5mm]
    \SubMatrix{.}{7-1}{9-last}{\}}[xshift=2.5mm]
    \SubMatrix{.}{10-1}{last-last}{\}}[xshift=2.5mm]
    \UnderBrace[yshift=3pt]{last-1}{last-last}{\va{a}}
\end{pNiceMatrix}
-\quad\begin{pNiceMatrix}[last-col]
\ev*{Q_0}_{1\eta_1}&\Block{3-1}{\quad \va{y}_{01\va{\eta}}}\\
\Vdots&\\
\Block[borders={bottom, tikz=dotted}]{1-1}{}\ev*{Q_0}_{1\eta_m}&\\
\ev*{Q_0}_{2\eta_1}&\Block{3-1}{\quad \va{y}_{02\va{\eta}}}\\
\Vdots&\\
\Block[borders={bottom, tikz=dashed}]{1-1}{}\ev*{Q_0}_{2\eta_m}&\\
\ev*{Q_1}_{1\eta_1}&\Block{3-1}{\quad \va{y}_{11\va{\eta}}}\\
\Vdots&\\
\Block[borders={bottom, tikz=dotted}]{1-1}{}\ev*{Q_1}_{1\eta_m}&\\
\ev*{Q_1}_{2\eta_1}&\Block{3-1}{\quad \va{y}_{12\va{\eta}}}\\
\Vdots&\\
\Block[borders={bottom, tikz=solid}]{1-1}{}\ev*{Q_1}_{2\eta_m}&\\
-\beta_{02}(0)\ev*{Q^\emptyset_0}_0&\Block{6-1}{\quad \va{g}}\\
-\beta_{02}(\frac{1}{2})\ev*{Q^\emptyset_0}_0&\\
\Block[borders={bottom, tikz=solid}]{1-1}{}-\beta_{02}(1)\ev*{Q^\emptyset_0}_0\\
\qty(V-\beta_{02}(0))\ev*{Q^\emptyset_1}_0&\\
-\beta_{02}(\frac{1}{2})\ev*{Q^\emptyset_1}_0&\\
-\beta_{02}(1)\ev*{Q^\emptyset_1}_0&
\CodeAfter
    \SubMatrix{.}{1-1}{3-last}{\}}[xshift=2.5mm]
    \SubMatrix{.}{4-1}{6-last}{\}}[xshift=2.5mm]
    \SubMatrix{.}{7-1}{9-last}{\}}[xshift=2.5mm]
    \SubMatrix{.}{10-1}{12-last}{\}}[xshift=2.5mm]
    \SubMatrix{.}{13-1}{18-last}{\}}[xshift=2.5mm]
    \UnderBrace[yshift=3pt]{last-1}{last-last}{\va{y}}
\end{pNiceMatrix} \;.
\end{equation}
\,\\
\end{widetext}


\bibliography{bio}

\end{document}